\documentclass[pdflatex,sn-mathphys-num]{sn-jnl}

\usepackage{graphicx}%
\usepackage{multirow}%
\usepackage{amsmath,amssymb,amsfonts}%
\usepackage{amsthm}%
\usepackage{mathrsfs}%
\usepackage[title]{appendix}%
\usepackage{xcolor}%
\usepackage{textcomp}%
\usepackage{manyfoot}%
\usepackage{booktabs}%
\usepackage{algorithm}%
\usepackage{algorithmicx}%
\usepackage{algpseudocode}%
\usepackage{listings}%
\usepackage{setspace}
\usepackage[colorinlistoftodos,prependcaption,textsize=tiny]{todonotes}
\usepackage{enumitem}

\theoremstyle{thmstyleone}%
\theoremstyle{thmstyletwo}%

\theoremstyle{thmstylethree}%
\newcommand{\xm}[1]{{\color{black}{#1}}} 

\begin{document}

\title[HelixFold-S1]{Reshaping Biomolecular Structure Prediction through Strategic Conformational Exploration with HelixFold-S1}


\author[1]{Lihang Liu} \equalcont{These authors contributed equally to this work.} 
\author[1]{Yang Liu} \equalcont{These authors contributed equally to this work.}
\author[1]{Xianbin Ye} \equalcont{These authors contributed equally to this work.}
\author[1]{Shanzhuo Zhang}
\author[1]{Yuxin Li}
\author[1]{Kunrui Zhu}
\author[1]{Yang Xue}
\author[1]{Jingbo Zhou}
\author[1]{Xiaonan Zhang}
\author*[1]{Xiaomin Fang} \email{fangxiaomin666@gmail.com}



\affil[1]{PaddleHelix Team, Baidu Inc}




\abstract{The abstract serves both as a general introduction to the topic and as a brief, non-technical summary of the main results and their implications. Authors are advised to check the author instructions for the journal they are submitting to for word limits and if structural elements like subheadings, citations, or equations are permitted.}


\abstract{
Generating large ensembles of candidate conformations is standard for improving biomolecular structure prediction. Yet aimless sampling is inefficient and costly, producing many redundant conformations with limited diversity, \xm{particularly for complex multimeric assemblies}. Here, we present HelixFold-S1, a guided planning approach \xm{specifically designed to enhance the structural prediction of biomolecular complexes} by strategically targeting the most informative regions of conformational space to produce accurate conformations. For each \xm{complex}, predicted inter-chain contact probabilities serve as a blueprint of the conformational space, guiding computational effort toward higher-probability, low-redundancy contacts that constrain structure generation. Across diverse biomolecular \xm{complex} benchmarks, HelixFold-S1 achieves markedly higher structural accuracy than traditional unguided methods while reducing sampling requirements by an order of magnitude. Predicted contact probabilities also provide a rough indicator of prediction difficulty and sampling utility. These results demonstrate that guided planning reshapes conformational exploration and enables more efficient and accurate structural inference.
}




\maketitle

\section*{Introduction}\label{sec1}
\begin{figure}[ht!]
    \centering
    \includegraphics[width=0.85\linewidth]{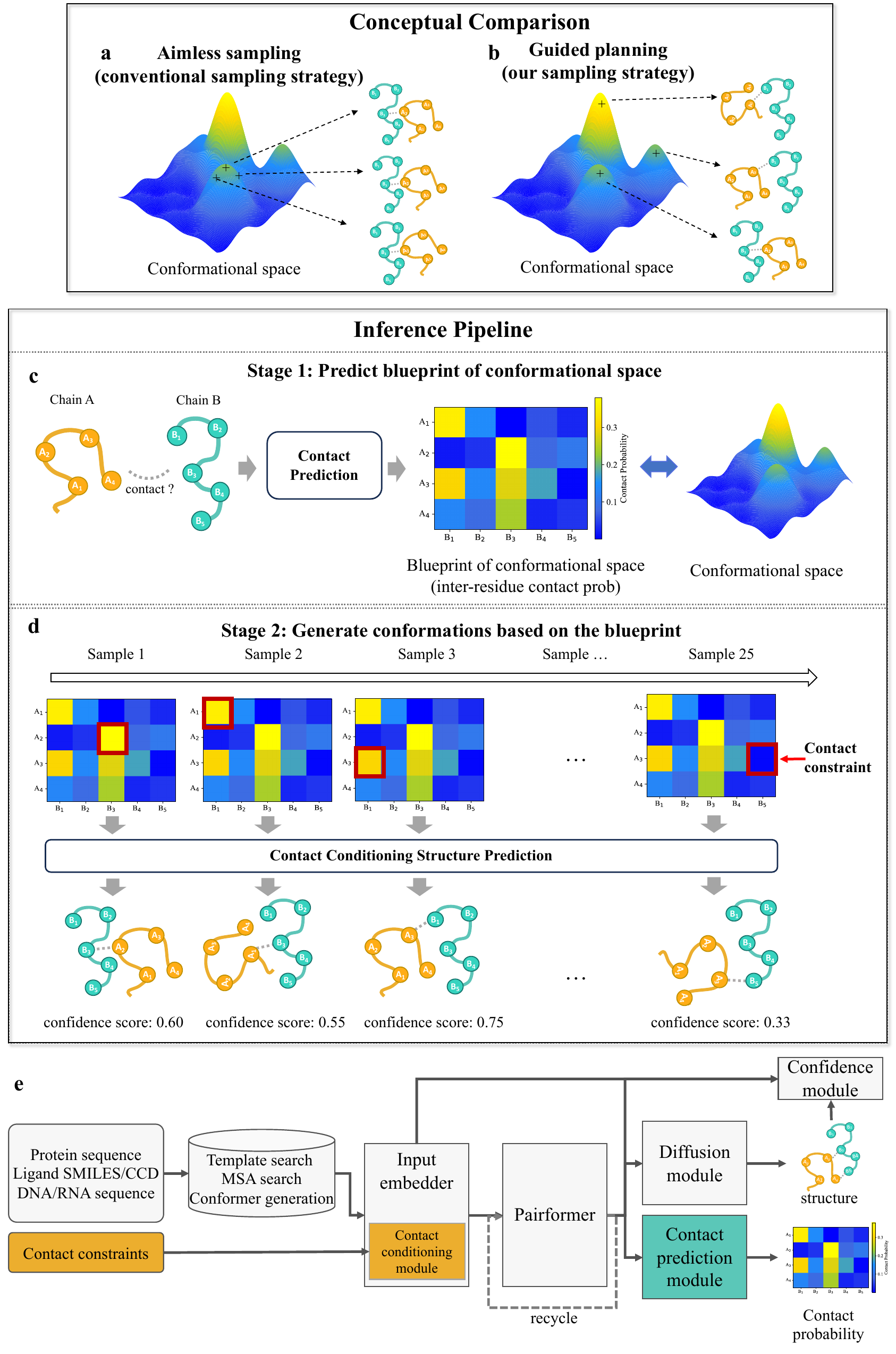}
    \caption{\textbf{Overall framework of HelixFold-S1.} \xm{\textbf{a, b,} Conceptual comparison of sampling paradigms: \textbf{a,} Aimless exploration in conventional conformation sampling, often concentrated within limited regions of the conformational space; \textbf{b,} Guided exploration in planning-based sampling, illustrating the strategic advantage of directing sampling toward diverse conformational states.} \textbf{c, d,} Two-stage inference of HelixFold-S1: \textbf{c,} stage 1 predicts a blueprint of the conformational space; \textbf{d,} stage 2 generates conformations based on this blueprint. \textbf{e,} Network architecture of HF-S1, which incorporates a contact conditioning module and a contact prediction module into the HelixFold3's model architecture.}
    \label{fig:architecture}
\end{figure}

Biomolecular structure prediction lies at the heart of computational biology, enabling progress in drug discovery, protein engineering, and the study of molecular interactions. Deep learning has transformed this field, with recent breakthroughs \cite{jumper2021highly,evans2021protein,abramson2024accurate,baek2021accurate,krishna2024generalized,lin2023evolutionary,fang2023method,fang2024helixfold,liu2024technical,mirdita2022colabfold,hayes2025simulating} exemplified by the AlphaFold \cite{jumper2021highly,evans2021protein,abramson2024accurate} and RoseTTAFold series \cite{baek2021accurate,krishna2024generalized}.
Despite these advances, accurate prediction of biomolecular complexes remains challenging.

\xm{To improve AlphaFold-like models of complexes, several specialized approaches have been proposed, such as denoising MSA profiles \cite{bryant2024improved}, integrating experimental constraints \cite{mirabello2024unmasking}, or optimizing interaction interfaces through architectural advancements \cite{bryant2022improved, akiyama2025scaling}. Despite this progress, achieving consistent accuracy across diverse biomolecular systems remains a significant challenge. In parallel, }improving the accuracy of a single prediction has proved difficult, and generating large ensembles of candidate conformations has therefore emerged as a simple yet powerful way to boost performance. The AlphaFold series improves precision through model ensembling, while AlphaFold3 \cite{abramson2024accurate} demonstrates that extensive sampling markedly enhances protein–antibody interface modeling. Recent studies \cite{wallner2023afsample,kalakoti2025afsample2,stein2022speach_af,yin2024evaluation} have generated thousands of candidates to expand conformational diversity. For instance, AFSample \cite{wallner2023afsample,kalakoti2025afsample2} introduces stochasticity via dropout and random template masking to improve accuracy. Other strategies \cite{xing2025af2claseq,monteiro2024high,wayment2024predicting,bryant2024structure} diversify predictions through repeated MSA sub-sampling to probe alternative structural states.

Although generating large ensembles of candidate structures can improve predictive accuracy, it comes at substantial computational cost. For instance, performing a thousand samplings may require an entire GPU day. The main reason such massive sampling is needed is that current sampling methods are largely unguided, leading to aimless exploration of conformational space. \xm{Specifically, predicting a biomolecular complex is essentially a search problem within a vast inter-chain conformational space, where only a few high-probability regions that correspond to potential binding modes are structurally plausible. The produced conformations often cluster around a single local region of the conformational landscape (Fig.~\ref{fig:architecture}a), producing redundant conformations that offer limited additional information. Consequently, sampling efficiency remains low.} It is often unclear if further sampling is beneficial, as simpler targets gain little from extensive ensembling, rendering much computation redundant. These challenges highlight the need for planning-driven strategies that can first illustrate the most potentially informative regions, and then intelligently navigate the conformational space to focus computational effort where it matters most, thereby achieving higher accuracy at far lower computational cost.

Here, we introduce HelixFold-S1, a guided sampling strategy \xm{designed for biomolecular complexes} that explores conformational space with exceptional efficiency and precision. \xm{In contrast to traditional stochastic search, HelixFold-S1 adopts a planning-based strategy to navigate the multi-modal landscape of complex structures. This approach strategically targets the most informative regions, which are conceptualized as high-probability modes within the conformational landscape, to facilitate the deliberate exploration of diverse and plausible binding modes (Fig. \ref{fig:architecture}b).} For each \xm{target complex}, we first predict inter-chain inter-residue contact probabilities. Interpreting these distributions as a coarse-grained structural blueprint of the conformational space provides a reduced representation that highlights where meaningful interaction patterns are most likely to occur (Fig.~\ref{fig:architecture}c). This structural map then guides the sampling, with computational effort concentrated in regions associated with higher-probability, low-redundancy contacts that act as spatial constraints during structure generation. The resulting conformations are evaluated and ranked by their confidence scores, enabling both accurate and resource-efficient modeling of complex biomolecular assemblies (Fig.~\ref{fig:architecture}d).

We evaluated HelixFold-S1 across diverse biomolecular interfaces, including protein–protein, protein–antibody, protein–ligand, protein–RNA, and protein–DNA. Compared with conventional unguided sampling, HelixFold-S1 substantially improves prediction accuracy at equivalent sampling effort, with the largest gains observed for challenging protein–antibody complexes. Remarkably, it achieves comparable accuracy to traditional methods using an order of magnitude fewer sampling steps. The approach is generalizable to other folding models to enhance structural precision. Predicted inter-chain contact probabilities correlate strongly with structural difficulty and provide a practical guide for allocating sampling effort across targets. Guided sampling also improves exploration of the conformational space, producing more diverse and higher-quality ensembles.

\section*{Results}
\subsection*{Guided Sampling Strategy and Architecture of HelixFold-S1}
The inference pipeline of HF-S1 comprises two stages. In stage 1, HF-S1 predicts a coarse blueprint of the conformational space (Fig.~\ref{fig:architecture}c) to identify the most critical inter-chain interaction sites. An additional Contact Prediction Module (CPM) is employed to estimate inter-chain inter-token contact probabilities based on the pair representation, representing the likelihood that any two tokens are in spatial proximity, i.e., any atom pair lies within 5\AA. This contact probability map provides a reduced, coarse-grained representation of the high-dimensional conformational space, serving as a structural blueprint to guide subsequent conformation sampling.
In stage 2, conformations are generated guided by the predicted blueprint. Contacts are prioritized according to their predicted probabilities and sequentially selected from the blueprint. Each selected contact is introduced as an additional constraint via the Contact Conditioning Module (CCM), which processes these constraints and generates structures that attempt to satisfy them. \xm{This design aligns with conditional structure prediction \cite{stahl2023protein,stahl2024modelling,heo2022multi}, where geometric priors bias the model toward specific states. Our CCM uses the inter-chain contacts from Stage 1 as constraints to effectively steer subsequent structure prediction.} By focusing on contacts with higher predicted probabilities, the model is more likely to produce multiple high-quality conformations that capture the most informative interactions. To reduce sampling redundancy, we employ a strategy called redundancy contact pruning (RCP). Once a conformation is generated using a particular contact, any contacts that are already satisfied within the resulting structure are excluded from further selection. All generated conformations are subsequently ranked according to model confidence scores, with the top-ranked prediction designated as the final output. \xm{Notably, RCP is an optional optimization; it enhances efficiency by sequentially pruning redundant contacts but can be bypassed for fully independent parallel generation. This allows HelixFold-S1 to balance high throughput via parallelism with refined, low-redundancy sampling, depending on available resources.}

The architecture of HelixFold-S1 (HF-S1) builds upon that of HelixFold3 (HF3) \cite{liu2024technical}, which reproduces the biomolecular structure prediction capabilities of AlphaFold3 (AF3) \cite{abramson2024accurate}, and further extends its capacity to more intelligently explore conformational space through the addition of two contact-centric modules: the CPM and the CCM (Fig.~\ref{fig:architecture}e). A contact is considered to exist between two tokens if any pair of atoms from the corresponding tokens lies within 5\AA. The CPM computes a contact probability matrix across all token pairs using the pair representations produced by the Pairformer, a specialized attention-based module designed to capture long-range pairwise token dependencies. The CCM, incorporated into the Input Embedder, enables the model to learn contact constraints and generate structures conditioned on these constraints. 

During inference, the two modules are employed sequentially: in the first stage (Fig.~\ref{fig:architecture}c), the CPM estimates the overall contact landscape, and in the second stage (Fig.~\ref{fig:architecture}d), the CCM generates conformations guided by the predicted contacts. \xm{In contrast, during the training process, the model is optimized within a multi-task framework that alternates between two primary objectives: the contact prediction task (utilizing the CPM) and the contact-conditioned structure generation task (utilizing the CCM). This dual-task optimization} allows the model to jointly learn inter-chain interactions and effectively leverage them during inference.

\subsection*{Improved Structural Accuracy across Complex Types}
\begin{figure}[ht!]
    \centering
    \includegraphics[width=1.0\linewidth]{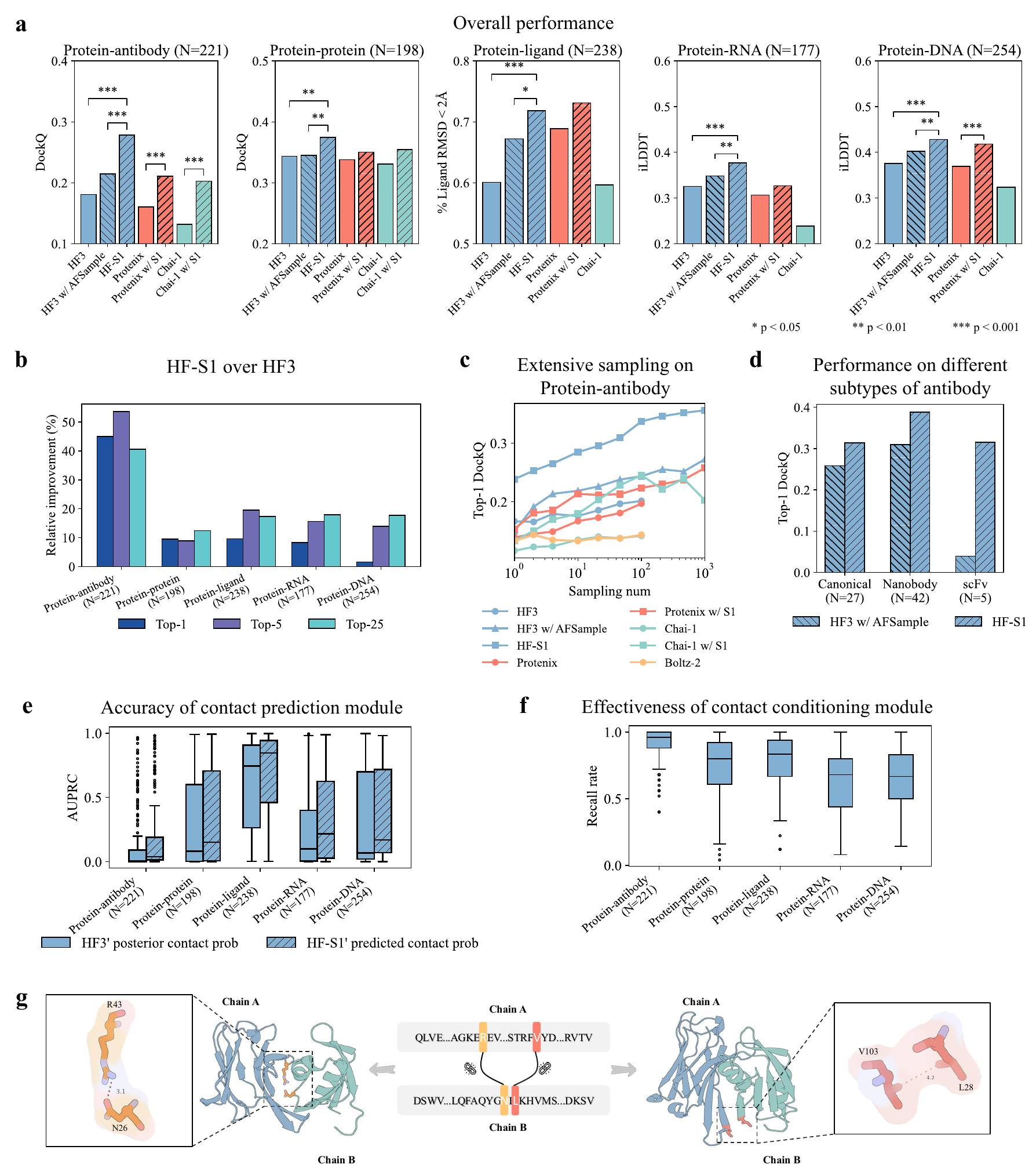}
    \caption{\textbf{Structural performance and module evaluation of HelixFold-S1 across multiple complex types.} \textbf{a,} Top-5 structural precision among 25 sampled conformations for various folding models. Benchmark complexes were collected from the RCSB PDB between January 1, 2022, and December 31, 2024, including protein–antibody ($n=221$), protein–protein ($n=198$), protein–ligand ($n=238$), protein–RNA ($n=177$), and protein–DNA ($n=254$) complexes. Statistical significance was determined using a two-sided paired t-test. On the protein-protein dataset, the difference between HF-S1 and HF3 was significant ($p\approx0.0029$); for all other datasets, the improvements were highly significant ($p<10^{-6}$). \textbf{b,} Relative improvements of HelixFold-S1 over HelixFold3 across different complex types in Top-1, Top-5, and Top-25 precision. \textbf{c,} Comparison of HF-S1 with various baseline models on protein–antibody complexes released in 2024 (n=74), showing that extensive sampling progressively improves structural accuracy. Some methods were sampled 1000 times, while others used 100 samples. \textbf{d,} Performance comparison between HF3 w/ AFSample and HF-S1 across different antibody types on the same 2024 protein–antibody complexes (n=74), using 1000 samples. \textbf{e,} Performance of the contact prediction module in HF-S1, evaluated using the area under the precision–recall curve (AUPRC). \textbf{f,} Effectiveness of the contact conditioning module, measured as the fraction of predicted structures satisfying contact constraints. In the box plots of \textbf{e} and \textbf{f}, the center line indicates the median; the box limits represent the 25th and 75th percentiles; and the whiskers extend to the maximum and minimum values, excluding outliers. \textbf{g,} Example structures predicted by HelixFold-S1 using two contact constraints (PDB ID: 8ozb).}
    \label{fig:overall}
\end{figure}

To systematically evaluate the performance of HF-S1, we constructed a test set comprising protein–antibody (n=221), protein–protein (n=198), protein–ligand (n=238), protein–RNA (n=177), and protein–DNA (n=254) complexes, collected between 2022.01.01 and 2024.12.31 from the RCSB PDB \cite{burley2017protein}.  To minimize overlap with the training data, all test samples were selected to have low sequence identity to the training set. Sequences were clustered by similarity \cite{steinegger2017mmseqs2}, and one representative per cluster was randomly chosen to ensure diversity and reduce redundancy. For protein–ligand complexes, any ligands that appeared in the training data were further excluded. 

We compared AF3-like models\footnote{AlphaFold3 is not available for commercial use and thus could not be tested. In addition, part of our benchmark overlaps with the Boltz-2 \cite{passaro2025boltz} training set, and Boltz-2 was therefore excluded.}, namely HF3 \cite{liu2024technical}, Protenix \cite{bytedance2025protenix}, and Chai-1 \cite{chai2024chai}, under different sampling strategies. For HF3, we examined two sampling-augmented variants: one following the AFSample strategy \cite{wallner2023afsample} (HF3 w/ AFSample) and another implementing our guided planning approach (HF-S1). \xm{To assess cross-model compatibility, we integrated HF-S1-predicted contacts into Protenix and Chai-1 (Protenix/Chai-1 w/ S1). This evaluation prioritizes sampling generalizability across architectures over absolute rankings. To manage the computational cost of benchmarking multiple external models, these variants were run in fully parallel mode without RCP; crucially, this trade-off does not compromise the fundamental assessment of our guided sampling strategy.} Because HF-S1 may employ a distinct definition of contacts from those folding models, the baseline models may not reach their best performance under this evaluation. Moreover, Chai-1 currently supports contact constraints only for protein–protein complexes, and results are therefore reported for protein–antibody and protein–protein systems only. For all methods, we generated 25 predicted structures per target and ranked them using each method’s confidence score.

Top-5 precision (Fig.~\ref{fig:overall}a) reveals that HF-S1 consistently outperforms both HF3 and HF3 w/ AFSample. Improvements are most significant for protein–antibody complexes, exceeding 55\% over HF3 and 33\% over HF3 w/ AFSample. This is followed by protein–protein and protein–RNA/DNA interfaces, while protein–ligand binding sites show the smallest gains. This likely reflects the structural diversity of antigen-binding sites, which makes conventional sampling less effective at capturing native-like geometries, whereas HF-S1’s guided sampling strategy increases the likelihood of generating correct interfacial contacts. For protein–protein and protein–RNA/DNA interfaces, HF-S1 provides moderate but clear improvements. By enforcing constraints on individual interfacial contacts, guided sampling helps capture correct local interactions, although single-contact constraints alone are not sufficient to fully determine the global geometry of these larger, more complex interfaces. Finally, protein–ligand binding sites are highly constrained, so while HF-S1 can still improve predictions, the relative gains are smaller compared with other complex types. Moreover, across different folding models, including HF3, Protenix, and Chai-1, incorporating the guided sampling strategy of HF-S1 leads to consistent performance gains, underscoring its broad effectiveness. In Fig.~\ref{fig:overall}b, we further demonstrate that the performance advantage of HF-S1 over HF3 is consistently observed across different Top-K precisions of the sampled conformations. The improvements are pronounced for Top-1, Top-5, and Top-25, underscoring HF-S1's robustness across multiple ranking thresholds, although the Top-1 gains exhibit greater variability, potentially reflecting sensitivity to the selection of conformations based on the confidence score.

Extensive sampling is particularly effective for protein–antibody interfaces \cite{bytedance2025protenix}, and HF-S1 shows the largest gains on these targets. To systematically evaluate the benefit of extensive sampling on these interfaces, we curated a dataset of 74 complexes released after 2024 and with residues less than 800 for efficiency, ensuring no overlap with Boltz-2’s training data \cite{passaro2025boltz}. Sampling was conducted with top-performing methods evaluated 1,000 times per complex, while others were sampled 100 times per complex. Top-1 precision on protein–antibody interfaces (Fig.~\ref{fig:overall}c) generally improves with increased sample size across all methods, particularly for those incorporating advanced strategies such as HF-S1 or AFsample. Importantly, HF-S1 achieves comparable precision to HF3 with AFsample using only 10 samples. This indicates that 10 HF-S1 samples are sufficient to reach the precision level that HF3 with AFsample requires 1,000 samples to achieve, representing just 1\% of the computational cost. To further assess performance across different antibody types, Fig.~\ref{fig:overall}d presents Top-1 precision for HF-S1 and HF3 with AFsample across canonical antibodies, nanobodies, and scFv based on the full 1,000-sample evaluation. HF-S1 demonstrates consistent improvements in Top-1 precision over HF3 w/ AFsample across all categories.

HF-S1 introduces two key components built on HF3: the Contact Prediction Module (CPM) and Contact Conditioning Module (CCM). To evaluate the CPM, we assessed the accuracy of its predicted contact probability matrices. For each target, token pairs corresponding to true contacts in experimental structures were treated as positives, and all others as negatives. We then computed the area under the precision–recall curve (AUPRC) between predicted and ground-truth contact maps, averaging across all targets. As a baseline, a posterior contact probability matrix for HF3 was derived by aggregating sampled conformations, with each entry defined as the inverse of the minimum inter-residue distance. Compared with HF3, HF-S1 consistently produced more accurate contact probabilities across diverse molecular types (Fig.~\ref{fig:overall}e). Protein–antibody complexes showed the lowest AUPRC, reflecting the difficulty of this prediction scenario, whereas protein–ligand complexes achieved the highest AUPRC, indicating a relatively simpler task. This aligns with Fig.~\ref{fig:overall}a: protein–antibody interfaces benefit most from guided sampling, while protein–ligand interfaces show smaller relative gains.
To examine the CCM, we evaluated whether HF-S1–predicted conformations adhered to specified contact constraints. We defined the contact satisfaction rate as the fraction of predicted structures in which the given contacts were realized. Across most test cases, HF-S1 achieved satisfaction rates above 70\% (Fig.~\ref{fig:overall}f), demonstrating its ability to effectively incorporate contact priors into structure prediction.

Fig.~\ref{fig:overall}g illustrates a representative case. Two inter-chain residue pairs (R43–N26 and V103–L28) were selected from HF-S1’s predicted contact map and applied as spatial restraints during folding, yielding two distinct binding conformations that both satisfied the defined contact distances (R43–N26 $\approx$ 3.1~\AA; V103–L28 $\approx$ 4.2~\AA).


\subsection*{Contact Probability as an Indicator of Prediction Difficulty and Sampling Utility}
\begin{figure}[ht!]
    \centering
    \includegraphics[width=\linewidth]{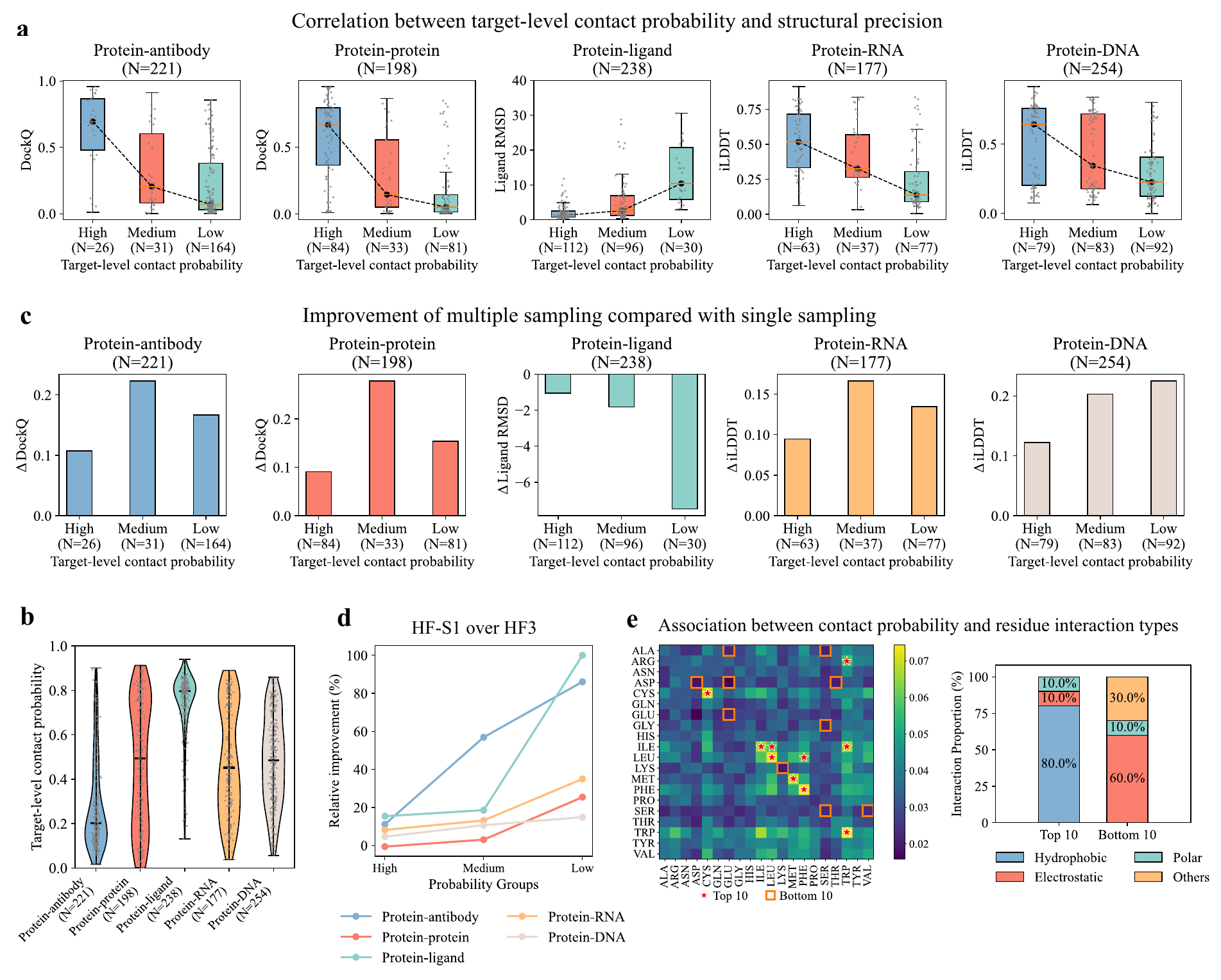}
    \caption{\textbf{Contact probability as an indicator of prediction difficulty and sampling utility.} \textbf{a,} Correlation between target-level contact probability (maximum value in the predicted matrix) and structural precision across interface types, evaluated using the Top-5 of 25 sampled conformations. In the box plots, the center line indicates the median; the box limits represent the 25th and 75th percentiles; and the whiskers extend to the minimum and maximum values (excluding outliers). \textbf{b,} Distribution of predicted contact probabilities across interface types, constructed from all benchmark targets in each category. \textbf{c,} Structural improvement from single to extensive sampling by HF-S1 across contact probability groups, measured using the best-performing conformations, showing larger gains for lower-probability interfaces. \textbf{d,} Relative Top-5 precision (of 25 sampled conformations) of HF-S1 compared with HF3 across contact probability levels, indicating stronger benefits on lower-probability targets. The sample numbers in this analysis match those in \textbf{a} and \textbf{c}. \textbf{e,} Predicted contact probabilities reflect residue physicochemical preferences.}
    \label{fig:contact_prob_analysis}
\end{figure}
The contact probability matrix predicted by HF-S1 serves a dual role: it not only informs the structural sampling strategy but also provides insight into the intrinsic difficulty of the structure prediction task, as well as the potential benefits of multiple sampling.

We first show that predicted contact probabilities can serve as a proxy for estimating target difficulty. Specifically, we defined the target-level contact probability as the maximum value in the predicted contact probability matrix for each target and analyzed its relationship with the Top-5 precision of HF-S1 using 25 samples (Fig.~\ref{fig:contact_prob_analysis}a). Targets were stratified into high, medium, and low groups, revealing a strong correlation across datasets: lower-probability targets consistently yielded poorer predictions, while higher-probability targets were predicted with high precision. This indicates that low contact probability generally corresponds to more difficult targets, which often require additional sampling to achieve accurate predictions. When grouped by complex type (Fig.~\ref{fig:contact_prob_analysis}b), a similar trend is observed: Protein–protein targets typically show higher target-level contact probabilities, while protein–antibody complexes cluster in the lower range, indicating greater structural uncertainty. Protein–ligand complexes consistently exhibit high contact probabilities, suggesting that HF-S1 can often localize ligand binding sites with high confidence. In contrast, protein–RNA and protein–DNA complexes display a broader distribution, reflecting greater variability in prediction difficulty across these categories.

We next examined whether target-level contact probabilities predict sampling utility. Targets were grouped based on their target-level predicted contact probability, and we analyzed the precision improvements of the best-performing multi-sample prediction relative to single-sample predictions (Fig.~\ref{fig:contact_prob_analysis}c). Targets with intermediate contact probabilities achieved the greatest improvements, while those with low or high probabilities saw more modest gains. Notably, although lower-probability targets did benefit from sampling, their improvements were generally smaller than those in the intermediate group. This trend is intuitive: for targets with high contact probabilities, accurate structures can often be recovered from a single sample, leaving limited room for further enhancement. For lower-probability targets, the predicted contact maps are weak across the board, suggesting that a much larger number of samples may be needed to identify accurate structures. In contrast, intermediate cases offer partial yet informative contact signals, enabling the model to better explore the structural landscape and refine its predictions through sampling. As protein–ligand complexes consistently exhibit high target-level contact probabilities, they did not follow this trend. 

We further examined performance across different contact probability groups by comparing HF-S1 (guided sampling) with HF3 (aimless sampling) (Fig.~\ref{fig:contact_prob_analysis}d). HF-S1 achieves the largest gains in the lower-probability group, while improvements are smaller in the higher-probability group. This trend suggests that targets with low contact probabilities are generally more challenging, and thus benefit more from guided exploration. In contrast, high-probability targets are easier to fold accurately, leaving less room for improvement.

Finally, we analyzed predicted contact probabilities across residue–residue pairs in the protein–protein dataset. Left panel of Fig.~\ref{fig:contact_prob_analysis}e: the averaged contact probability matrix across all residue pairs, with the pairs showing the highest probabilities indicated and those with the lowest probabilities highlighted. Right panel of Fig.~\ref{fig:contact_prob_analysis}e: compositional differences between the top 10 and bottom 10 pairs ranked by mean probability. Hydrophobic pairs such as Leu–Leu and Met-Met tend to show higher predicted probabilities, which could reflect their relatively frequent occurrence in tightly packed, energetically favorable environments. Conversely, polar or charged residues, such as Asp, Glu, and Lys, often exhibit lower probabilities, perhaps due to their general solvent exposure or context‐specific interaction tendencies. These patterns indicate that contact prediction models may capture fundamental physicochemical preferences underlying residue interactions \cite{ppi_analysis2008,ppi_analysis2013,ppi_analysis2018,ppi_analysis2021}.

\subsection*{Guided Sampling Improves Exploration of Conformation Space}
\begin{figure}[ht!]
    \centering
    \includegraphics[width=\linewidth]{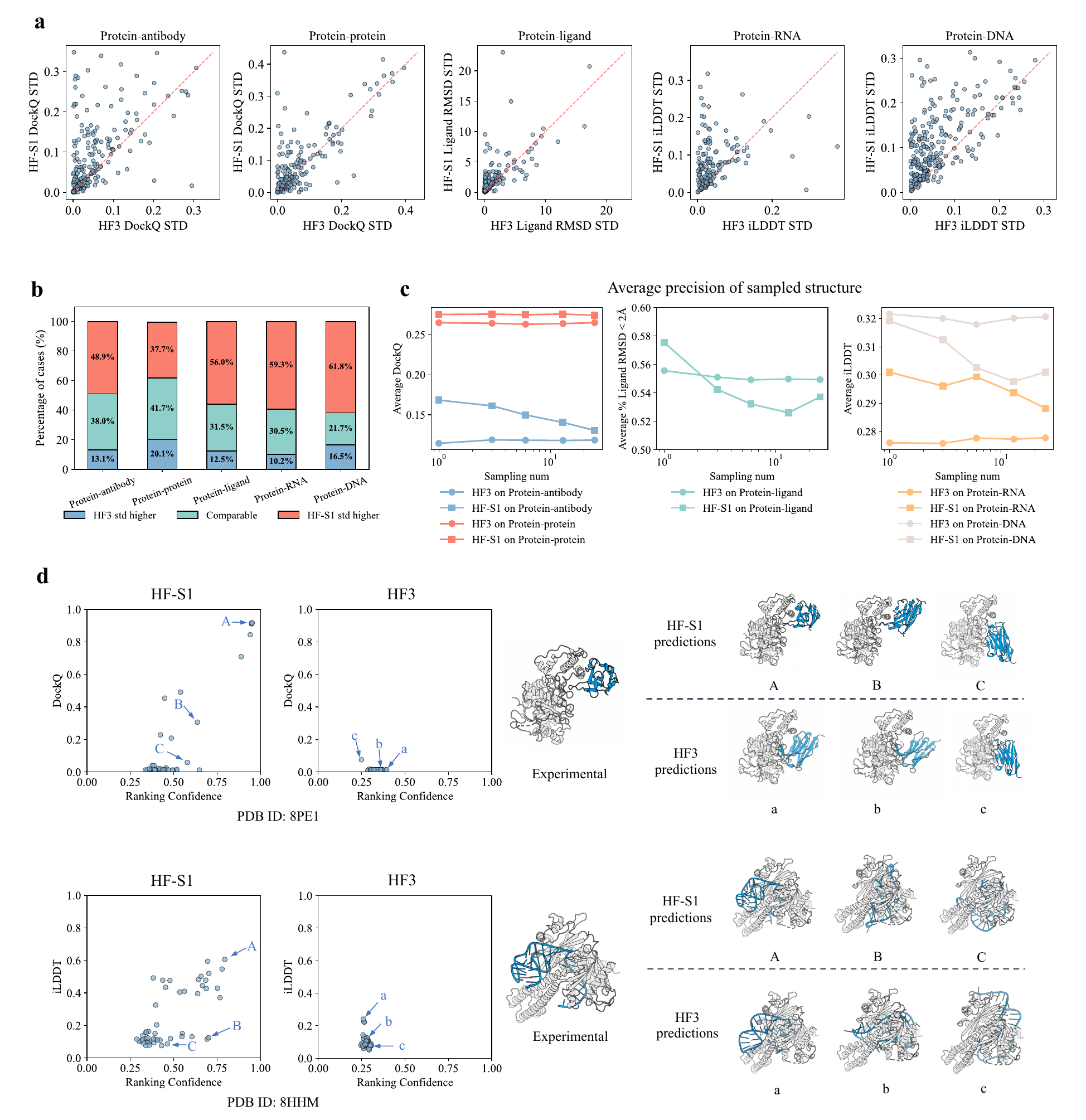}
    \caption{\textbf{Guided sampling improves exploration of conformational space.} \textbf{a,} Target-level standard deviation of structural precision among sampled conformations for HF3 and HF-S1. Each point represents a target; higher values indicate greater conformational diversity. \textbf{b,} Fraction of targets with higher standard deviation for HF-S1, higher for HF3, or comparable, summarized across different interface types.\textbf{c,} Structural precision as a function of the number of sampled conformations. \textbf{d,} Distribution of ranking confidence and structural precision of sampled structures for two representative complexes (PDB: 8PE1 and 8HHM). HF-S1 shows a broader distribution of structural precision, whereas HF3 predictions are more concentrated. Right panels show the corresponding predicted structures by HF-S1 and HF3 compared to experimental structures.}
    \label{fig:conformation_diversity}
\end{figure}

To test if guided sampling improves conformational exploration, we compared HF-S1 (guided planning) with HF3 (aimless sampling). We calculated the standard deviation (std) of precision scores per target as a proxy for structural diversity (Fig.~\ref{fig:conformation_diversity}a), where higher values indicate a broader exploration of the landscape. HF-S1 generally shows higher std values, with data concentrated in the upper-left region of the plot, reflecting more diverse ensembles. Targets were then grouped based on the difference in std between HF-S1 and HF3 (Fig.~\ref{fig:conformation_diversity}b): for protein–ligand complexes, a difference $\ge1$ indicates HF-S1 std higher, $\leq-1$ indicates HF3 std higher, and intermediate values are comparable; for all other complex types, the threshold is 0.1. Across complex types, most targets fall into the HF-S1 std higher category, particularly protein–antibody, protein–RNA, and protein–DNA complexes, where the HF-S1 std higher category accounts for roughly twice as many targets as the HF3 std higher category. This enhanced diversity suggests that guided planning effectively directs sampling toward multiple plausible interaction geometries.

To examine how guided planning shapes the sampling trajectory, we tracked the average precision of all conformations accumulated up to each sampling step $K$ (Fig.~\ref{fig:conformation_diversity}c). For HF3, precision remains nearly constant across steps, as the order of sampled conformations is effectively random. In contrast, HF-S1 exhibits a gradual decline in precision as sampling proceeds, reflecting its design to explore progressively less probable contact configurations. The higher precision observed in the early sampling stages arises from the use of higher-probability contacts as structural constraints, which guide the generation of more accurate conformations. Notably, HF-S1’s initial conformations already outperform HF3 across most interface types, excluding protein–DNA complexes. This demonstrates that leveraging contact probabilities as structural constraints not only improves efficiency but also prioritizes structurally meaningful regions of the conformational landscape.

We compared the aimless sampling of HF3 with the guided strategy of HF-S1 using two representative examples (Fig. \ref{fig:conformation_diversity}d): the Gel4–Nb4 nanobody complex (PDB ID: 8pe1) \cite{redrado2024broad} and the Cas12m2–crRNA protein-RNA heterodimer (PDB ID: 8hhm) \cite{omura2023mechanistic}. Analysis of predicted confidence and interface precision shows that HF-S1 consistently generates a broader spectrum of structures. These samples span a wide range of confidence and precision values, including a higher fraction of high-accuracy predictions. In contrast, HF3 samples cluster within narrower ranges, indicating more limited structural diversity and fewer high-quality candidates.

\section*{Discussion}
Extensive sampling is standard for structural accuracy, yet unguided methods are computationally inefficient and lack insight into target difficulty. We propose that intelligent exploration of the conformational landscape enables efficient resource allocation. By directing sampling toward informative regions, our approach enhances both efficiency and accuracy. Furthermore, predicted contact probabilities serve as a proxy for target difficulty, guiding users on the required sampling scale for convergence.

\xm{Beyond efficiency, this Search-then-Filter paradigm avoids local minima and consistently achieves superior Top-1 accuracy. For challenging targets, such as multi-epitope antibodies, cryptic pockets, or transitioning GPCRs, this prioritized ensemble ensures rare but biologically active conformations are not overlooked, providing value far beyond a single structure.}

Despite these advances, assessing structural confidence remains challenging. Current metrics frequently misidentify the most accurate structures. High-confidence predictions may lack accuracy, while low-confidence sets often contain near-native conformations (Supplementary Figure 3). Refining these via sophisticated scoring or ensemble-based calibration could optimize candidate selection and prevent overlooking high-quality structures. Moreover, guided planning can exhibit diminishing returns. Greedy contact selection from highest to lowest probability may progressively sample less informative regions (Fig.~\ref{fig:conformation_diversity}c). Implementing early-stopping based on contact saturation or plateauing structural quality could sustain sampling efficiency. Finally, maintaining diversity is critical. Current sequential contact exclusion reduces redundancy but remains computationally serial. Although this strategy is effective (Supplementary Figure 4), more proactive methods are needed to preserve diversity more efficiently.

Intelligently navigating the conformational space remains a critical challenge. Sophisticated strategies, such as Monte Carlo Tree Search or reinforcement learning, could dynamically balance exploration and exploitation. Such methods promise to identify high-quality, diverse conformations more efficiently, enhancing the robustness of structure prediction across complex biomolecular targets.

\section*{Methods}\label{sec11}

\subsection*{Model Architecture}
HF-S1 builds upon the HF3 architecture and is designed to support two complementary tasks: inter-chain contact prediction and contact-conditioned structure prediction. To this end, HF-S1 introduces two additional components: the Contact Prediction Module and the Contact Conditioning Module, which extend the base architecture to enable contact-level reasoning and constraint-based structure generation, respectively.

The contact prediction task aims to estimate the inter-chain contact distribution of a given protein complex. Various input features—including sequence, multiple sequence alignment (MSA), and template information—are first encoded and then processed by a Pairformer module to generate single and pair representations. These pairwise representations are subsequently passed to a dedicated Contact Prediction Module (CPM), which includes a Pairformer Stack and performs a binary classification task on each element of the pairwise representation to output contact probabilities for each inter-chain token pair. A contact is defined as the presence of any atom pair between two tokens within 5\AA~in 3D space. The resulting pairwise contact probability matrix captures the inter-chain contact distribution of the complex and serves as an informative intermediate representation. By operating in the simplified space of contacts—rather than directly in the complex, high-dimensional structural space—the model achieves greater computational efficiency and facilitates the contact-conditioned structure prediction task.


The contact-conditioned structure prediction task introduces a Contact Conditioning Module (CCM) to incorporate external contact constraints. These constraints are represented as a binary matrix $\{c_{ij}\}$, where $c_{ij} \in {0, 1}$ indicates whether token pair $(i, j)$ is in contact (1 if any atom pair is within 5\AA, and 0 otherwise). This matrix is projected through a linear layer and then fused into the pairwise activations within the Input Embedder of the model. During training, for each complex, 0–10 inter-chain contacts are randomly sampled from the contact set extracted from its ground-truth structure and provided as input. During inference, contact constraints are sampled from the contact probability matrix produced by the contact prediction task; the specific sampling strategy is described in a later section. The model learns to utilize the provided contact information to enhance structure prediction accuracy.

\subsection*{Training Regime}
Parameters of the newly introduced Contact Prediction Module and Contact Conditioning Module are randomly initialized, while the remaining parts of the model inherit weights from the pre-trained HF3. Fine-tuning is performed using the same training dataset as HelixFold3, which includes Protein Data Bank (PDB) \cite{burley2017protein}  structures released before September 30, 2021, supplemented with self-distillation data to enhance generalization. The training follows a three-stage fine-tuning strategy: the first stage focuses on the contact-conditioned structure prediction task to improve complex structure accuracy with inter-chain contact constraints; the second stage adds the contact prediction task, jointly optimizing the model for both tasks; the third stage follows the same setting of the second stage but extends to larger crop size.

In the first \xm{fine-tuning} stage, the model is trained exclusively on the contact-conditioned structure prediction task. For each training sample, inter-chain contacts are extracted from experimentally determined complex structures to form a ground-truth contact set $\mathcal{C}$. This set contains all inter-chain token pairs where at least one atom from each token lies within 5\AA~in three-dimensional space.  Contact conditioning is applied with 70\% probability: 1–10 token pairs are uniformly sampled from $\mathcal{C}$ and provided as binary contact constraints. In the remaining 30\% of samples, no contact constraints are used, which helps maintain the model’s ability to predict structures without external guidance. 
\xm{While trained on individual experimental structures, the CPM learns underlying interaction probabilities by aggregating structural patterns from the vast diversity of homologous complexes and motifs in the training set. This allows the module to map a protein pair to a multi-modal probability landscape during inference, representing a global statistical distribution of plausible binding modes rather than a single deterministic state.}

\xm{In the second and third fine-tuning stages, the model is optimized within a multi-task framework that simultaneously addresses both the contact-conditioned structure prediction task and the contact prediction task. The latter} aims to estimate the probability that each inter-chain token pair is in atomic contact, serving as a basis for generating contact constraints during inference. \xm{During the specific optimization of the contact prediction task in these stages, the Contact Conditioning Module (CCM) is intentionally not activated; this is a deliberate design to ensure the CPM learns to infer intrinsic interaction patterns directly from raw features (MSA and sequences) without the "leakage" of ground-truth constraints.} To supervise this task, a binary classification loss is applied over all inter-chain token pairs:
$$
\mathcal{L}_{\text{contact}} = \frac{1}{|\mathcal{P}|} \sum_{(i,j) \in \mathcal{P}} \text{cross\_entropy}(p_{ij}^{contact}, y_{ij}^{contact}).
$$
Here, $\mathcal{P}$ denotes the set of all token pairs $(i, j)$ such that token $i$ and token $j$ belong to different chains. $p_{ij}^{contact}$ is the predicted probability of contact between tokens $i$ and $j$. $y_{ij}^{contact} = 1$ if $(i,j) \in \mathcal{C}$ (i.e., the token pair is in contact), and $y_{ij}^{contact} = 0$ otherwise. During training, half of the samples are used for the contact-conditioned structure prediction task, following the protocol established in the first stage, while the other half are dedicated to training the contact prediction task, which guides the model to estimate inter-chain contact probability distributions.

The loss function largely follows the original AF3/HF3 formulation, with an additional contact loss term introduced during fine-tuning:
\[
\mathcal{L}_{\text{loss}} = \alpha_{\text{confidence}} \mathcal{L}_{\text{confidence}} + \alpha_{\text{diffusion}} \mathcal{L}_{\text{diffusion}} + \alpha_{\text{distogram}} \mathcal{L}_{\text{distogram}} + \alpha_{\text{contact}}\mathcal{L}_{\text{contact}},
\]
with hyperparameters $\alpha_{\text{confidence}}=0.01$, $\alpha_{\text{diffusion}}=4$, and $\alpha_{\text{distogram}}=0.3$. The contact loss coefficient $\alpha_{\text{contact}}$ is set to 1 during training samples used for the contact prediction task and 0 during samples used for the contact-conditioned structure prediction task. The definitions of all other loss terms remain consistent with those in AF3.

All stages use the Adam optimizer \cite{kingma2014adam} with parameters $\beta_1 = 0.9$, $\beta_2 = 0.95$, and $\epsilon = 10^{-8}$, and a learning rate of $2 \times 10^{-4}$. The mini-batch size is fixed at 128 for all stages. The first fine-tuning stage consists of 10,000 training steps with a crop size of 384. The second fine-tuning stage extends to 20,000 steps, also with a crop size of 384. The third stage continues training for an additional 3,000 steps with an increased crop size of 640. We conducted model training on 128 Nvidia A100 GPUs, with a total duration of approximately 10 days.

\subsection*{Inference Regime}
The inference process of HF-S1 (illustrated in Fig.~\ref{fig:architecture}a) consists of three stages: Contact Prediction, Contact Sampling, and Contact-Guided Structure Prediction and Ranking.

In the Contact Prediction stage, the contact prediction task is executed five times to reduce prediction variance, producing five contact probability matrices. These matrices are averaged element-wise to generate the final contact probability matrix, where each element $p_{ij}^{contact}$ represents the predicted contact probability between tokens $i$ and $j$. Only inter-chain contact probabilities are retained, with intra-chain contacts set to zero. For protein–antibody complexes, contact sampling is performed exclusively between the antigen chain and each antibody chain (heavy and light), excluding contacts between heavy and light chains.

In the Contact Sampling stage, inter-chain contacts are selected sequentially in descending order according to their predicted contact probabilities. Each selected contact is used as a binary constraint in the subsequent structure prediction step to generate diverse candidate structures. To improve sampling efficiency and avoid redundancy, two strategies are adopted: redundant contact pruning and enriched sampling of previously identified contact sets.
We denote the sets of contacts extracted from previously predicted structures as $C_1, C_2, \ldots$, where each $C_k$ corresponds to the contacts obtained from the $k$-th predicted structure, following the same ground-truth extraction method described earlier.
During sampling, redundant contact pruning excludes any candidate contact that overlaps with contacts already present in the union of all previously sampled sets $\bigcup_{i=1}^{k-1} C_i$. Here, overlapping means the candidate contact appears in any previously extracted contact set. This ensures that each newly sampled contact introduces novel constraints and helps maintain diversity among sampled structures.
As sampling progresses, the predicted contact probabilities of remaining candidates naturally decrease. When these probabilities fall below a threshold (set as $0.2 \cdot \max_{i,j} p_{ij}^{contact}$), the benefit of exploring new low-confidence contacts diminishes. At this point, instead of sampling new contacts, the algorithm enriches sampling by iterating through the existing contact sets $C_1, C_2, \ldots$ in order. Contacts are drawn from these sets cyclically to further exploit high-confidence information until the total sampling budget $S$ is reached.

In the Contact-Guided Structure Prediction and Ranking stage, each sampled contact is treated as a binary constraint and passed into the contact-conditioned structure prediction task to generate a candidate structure. A confidence score, named $\text{ranking\_confidence}$, is computed for each structure, and the final prediction is selected as the one with the highest confidence among all candidates. Drawing inspiration from the AF3, we define the confidence score as a weighted average of the pTM and ipTM scores, with an additional penalty term for structural clashes. The score is computed as follows:
\[
\text{ranking\_confidence} = 0.2 \cdot \text{pTM} + 0.8 \cdot \text{ipTM} - 1.0 \cdot \text{has\_clash},
\]
where $\text{pTM}$ represents the predicted TM-Score for the full complex, indicating the confidence for overall structural accuracy. $\text{ipTM}$ represents the interface predicted TM-Score for the full complex, focusing on the accuracy of interfacial interactions. $\text{has\_clash}$ is a binary term indicating the presence of obvious clashes between polymer chains in the predicted structure. Detailed definitions of $\text{pTM}$, $\text{ipTM}$, and $\text{has\_clash}$ can be found in the AF3 paper \cite{abramson2024accurate}.

We adopt consistent inference settings across structure prediction tasks, including our method and the baselines Boltz‑2\cite{passaro2025boltz}, Protenix\cite{bytedance2025protenix}, and Chai‑1\cite{chai2024chai}. Each prediction is refined using 10 recycling iterations and 200 diffusion steps, where the diffusion module is run once to generate a single structure per input. In the corresponding figures, lines represent these average outcomes, while shaded areas indicate the variability between the two runs. Notably, the inference configuration for HF3 w/ AFsample adopts a more sophisticated multi-setting approach, according to the AlphaFold settings used in AFsample \cite{wallner2023afsample}. The complete inference specifications for HF3 with AFsample integrate three distinct hyperparameter settings as detailed in Table \ref{tab:af_settings}.
Protenix w/ S1, Chai-1 w/ S1, and Boltz-2 w/ S1 follow the same workflow as HF-S1, differing only in the backbone structure prediction module employed during the final stage. Contact probability outputs from HF-S1 are ranked from highest to lowest, and the corresponding contact pairs are incorporated as external geometric constraints to guide structure generation. It should be noted that Chai-1 w/ S1 is restricted to protein–protein/antibody datasets, as it only accepts residue-residue contact constraints as input. All methods construct MSAs using their respective built-in sequence search tools.

\begin{table}
\centering
\begin{tabular}{lcccc}
\toprule
Settings & Templates & Dropout & Recycles & Ratio \% \\
\midrule
setting-1 & Yes & Yes & 3 & 30 \\
setting-2 & No & Yes & 3 & 30 \\
setting-3 & No & Yes & 9 & 40 \\
\bottomrule
\end{tabular}
\caption{Inference configurations of HF3 w/ AFsample. The term \textit{Templates} indicates whether structural templates were employed. \textit{Dropout} denotes whether the dropout mechanism was activated. \textit{Recycles} signifies the number of recycling operations utilized, with a default value of 3. \textit{Ratio} represents the proportion that this particular setting occupies within the entire sampling process.}
\label{tab:af_settings}
\end{table}

\subsection*{Evaluation Data}
Evaluation sets for protein–protein, protein–ligand, protein–RNA, and protein–DNA interfaces were constructed from all PDB entries released between May 1, 2022 and December 31, 2024, with each structure expanded to Biological Assembly 1. Interfaces were defined as pairs of entities with a minimum heavy-atom distance below 5 Å. Protein–antibody complexes were sourced from SAbDab \cite{schneider_sabdab_2021} within the same date range, using symmetric units instead of Biological Assembly 1.

For targets collected from the PDB, complexes with resolution worse than 4.5 Å or exceeding 1400 tokens under our tokenization scheme were removed. Polymer–polymer interfaces were excluded if both polymers shared more than 40\% sequence identity with two chains from the same PDB entry in the training set. For protein–ligand interfaces, the following criteria were applied: (1) only ligands with CCD codes absent from the training set were retained; (2) covalently bound ligands, including those involved in glycosylation, were excluded; (3) ligands containing five or fewer atoms or occurring in ten or more PDB entries were removed; (4) only ligands with molecular weights between 100 and 900 Da were retained; (5) ligands were required to exhibit a \textit{ranking\_model\_fit} score of at least 0.5, as reported in the RCSB structure validation dataset, indicating above-median model quality for X-ray crystallographic structures \cite{berman_protein_2000}; and (6) binding pockets were required to include between 5 and 100 protein residues within 5 Å of the ligand. 

We clustered the remaining targets by grouping proteins with nine or more residues at 40\% sequence identity, while nucleic acids and proteins with nine or fewer residues were clustered at 100\%, using MMseqs2 with a minimum coverage of 80\% and default clustering mode. Each interface was assigned a binary, order-independent cluster ID based on entity pairs—(polymer1\_cluster, polymer2\_cluster) for polymer–polymer interfaces and (polymer\_cluster, ligand\_CCD-code) for protein–ligand interfaces. Evaluation was performed on one representative entry per cluster.

Protein–antibody complexes were sourced from SAbDab \cite{schneider_sabdab_2021}, including only those with resolution better than 9 Å, and containing antigen chains. The antigen sequences were grouped into clusters based on a 40\% sequence identity threshold. Subsequently, we retained only those clusters that contained no cases released prior to September 30, 2021. Further filtering was applied to select cases released within the period from May 1, 2022, to December 31, 2024. Ultimately, one case was chosen from each of the remaining clusters to constitute the evaluation set for protein-antibody analysis. For efficiency considerations in the extensive sampling test(as shown in Fig.~\ref{fig:overall}c), we exclusively selected samples released between January 1, 2024, and December 31, 2024, and filtered out those with over 800 residues, ultimately retaining 74 samples.

\subsection*{Evaluation Metrics}
To evaluate structure prediction performance across different interaction types, we adopt distinct metrics tailored to the characteristics of each molecular interface.

Protein–protein complexes, including protein–antibody interactions, are evaluated using DockQ \cite{dockq_v1}, which integrates interface RMSD, FNAT, and FNAS to provide a reliable summary of interface quality. For protein–antibody complexes specifically, all antibody chains are treated collectively as the “ligand”, and DockQ is computed over the interface between the antibody and the rest of the complex using the DockQ v1 implementation.

Nucleic acid–protein interfaces, including both protein–RNA and protein–DNA complexes, are assessed using interface LDDT (iLDDT) \cite{btt473_lddt}, computed over atom pairs across different chains within a 30\AA~inclusion radius to accommodate the larger and more diffuse interaction footprints characteristic of nucleic acids.

Protein–ligand complexes are evaluated using pocket-aligned RMSD, which measures ligand pose accuracy after aligning the predicted structure to the binding pocket of the ground truth. The pocket is defined as all heavy atoms within 10\AA~of any heavy atom of the ligand in the ground truth structure, restricted to the primary protein chain—identified as the chain containing the most atoms within this radius. The $C^\alpha$ atoms of this pocket are used to perform a least-squares alignment between predicted and reference structures. After alignment, a symmetry-corrected ligand RMSD is computed over all heavy atoms of the ligand using RDKit’s Chem.rdMolAlign.CalcRMS \cite{rdkit}, which aligns the ligands while accounting for molecular symmetry before computing the final deviation.

The precision of the Top-K ranked structures is defined as the highest accuracy achieved among the top K structures ordered by their confidence score.

\section*{Data Availability}
To train HelixFold-S1, PDB can be downloaded at https://www.rcsb.org/docs/programmatic-access/file-download-services and AlphaFold Protein Structure Database as the distillation dataset can be downloaded at https://ftp.ebi.ac.uk/pub/databases/alphafold/v2/. The test set are filtered and clustered from PDB with conditions detailed in Methods. The protein-antibody complexes for test can be downloaded at https://opig.stats.ox.ac.uk/webapps/sabdab-sabpred/sabdab/. Detailed data processing procedures, including filtering, clustering, and dataset construction, are described in the Methods section.

\section*{Code Availability}
The source code, trained weights, and inference scripts for HelixFold-S1 will be made publicly available upon acceptance. During the peer review period, the code repository is accessible to reviewers at https://anonymous.4open.science/r/HelixFold-S1-D37C/README.md. In addition, a web service for HelixFold-S1 is accessible at https://paddlehelix.baidu.com/app/all/helixfold3/forecast, providing efficient and accurate structure predictions.

\section*{Acknowledgments}
This work was supported by the National Science and Technology Major Project (2023ZD0120803).

\section*{}

\clearpage
\bibliography{sn-bibliography}

\end{document}